\documentstyle[aaspp4]{article}
\input epsf.sty
\def\jepsfbox#1{\typeout{#1} \epsfbox{#1}}
\def\plotone#1{\centering \leavevmode
\epsfxsize=\columnwidth \jepsfbox{#1}}
\def\unsetyr{\def\oyear{\relax}\def\cyear{\relax}}
\def\setyr{\def\oyear{(}\def\cyear{)}}
\unsetyr
\def\jcite#1{\setyr\cite{#1}\unsetyr}
\def\rmmat#1{{\hbox{\rm #1}}}
\def\rmscr#1{\rmmat{\scriptsize #1}}
\newcommand{\be}{\begin{equation}}
\newcommand{\ee}{\end{equation}}
\newcommand{\ba}{\begin{eqnarray}}
\newcommand{\ea}{\end{eqnarray}}
\newcommand{\ie}{{\it i.e.~}}

\def\eqref#1{Equation~\ref{eq:#1}}
\def\figref#1{Figure~\ref{fig:#1}}

\begin{document}
\newcommand{\bfi}{{\bf B}} \newcommand{\efi}{{\bf E}}
\newcommand{\lel}{{\lambda_e^{\!\!\!\!-}}}
\newcommand{\me}{m_e}
\newcommand{\mcs}{{m_e c^2}}
\title{What is the nature of RX~J0720.4-3125?}
\author{Jeremy S. Heyl \\ jsheyl@ucolick.org}
\author{Lars Hernquist\altaffilmark{1} \\ lars@ucolick.org}
\affil{Lick Observatory,
University of California, Santa Cruz, California 95064, USA}
\altaffiltext{1}{Presidential Faculty Fellow}

\begin{abstract}
RX~J0720.4-3125 has recently been identified as a pulsating soft X-ray source 
in the ROSAT all-sky survey with a period of 8.391 s.  Its spectrum is
well characterized by a black-body with a temperature of $8 \times
10^5$ K.  We propose that the radiation from this object is thermal
emission from a cooling neutron star.  For this black-body temperature
we can obtain a robust estimate of the object's age of $\sim 3 \times
10^5$ yr, yielding a polar field $\sim 10^{14}$ G for magnetic-dipole
spin down and a value of ${\dot P}$ compatible with
current observations.
\end{abstract}

\section{Introduction}

``Breaking'' (\cite{Mere95}) or ``anomalous'' (\cite{vanP95}) x-ray
pulsars (AXPs) typically have pulsed X-ray emission with steadily
increasing periods $\sim$ 10 s, X-ray luminosities $\sim
10^{35}-10^{36}$ erg/s, soft spectra, and no detected companions or
accretion disks.  \jcite{Habe97} have recently identified
RX~J0720.4-3125 as a pulsating, soft X-ray source with a period of
8.391 s.  They estimate its bolometric luminosity to be $2.6 \times
10^{31} d_{100}^2$ erg/s, where $d_{100}$ is the object's distance in
units of 100 pc.

\jcite{Wang97} has proposed that accretion onto a weakly magnetized
neutron star powers RX~J0720.4-3125, and that for the neutron star to
spin down to 8.391 s within a Hubble time, it either must have been born
with a period $\sim 0.5$ s or have experienced magnetic field decay.
Our proposal begins with the second footnote of \jcite{Wang97} which
suggests the possibility that RX~J0720.4-3125 is powered by an internal
heat source.  We propose that the neutron-star cooling powers
RX~J0720.4-3125.  This would give it an age $\sim 3 \times 10^5$ yr,
much younger than the Wang's (1997) estimated age $\gtrsim 10^9$ yr.

\section{Calculations}

In a previous paper (\cite{Heyl97c}), we have argued that AXPs may be
powered by neutron star cooling through an accreted envelope.  Here we
examine the cooling evolution of neutron stars with polar field
strengths ranging from 0 to $10^{16}$ G.  The zero field calculation is
based on the envelope model of \jcite{Hern84b}.  The results of
\jcite{Hern85} describe the model envelope for $10^{13}$ G, and the
ultramagnetized cases ($10^{14}-10^{16}$ G) are calculated from
the models of \jcite{Heyl97a}. 

We choose a simple cooling model with the modified URCA process
(\cite{Shap83}) and photon cooling competing (\cite{Heyl97b}).  Both the
envelope calculations and the cooling calculations are performed in the
frame of the surface; consequently, the observed effective temperature
for a given core temperature is proportional to $g_s^{1/4}$ ($g_s$ is
the gravitation acceleration at the surface) and $(1+z_g)^{-1}$ where
\be
(1+z_g)^{-1} = \sqrt{1 - \frac{2 G M}{R c^2}}.
\ee
To obtain the effective temperature as observed at infinity 
for a given equation of state (\ie $M$ and $R$), these two corrections
must be performed.  Additionally, since for a magnetized neutron star,
the thermal emission is not isotropic along the surface, we present the
flux-weighted mean ${\bar T}_\rmscr{eff}$ over the entire surface. 

\begin{figure} 
\plotone{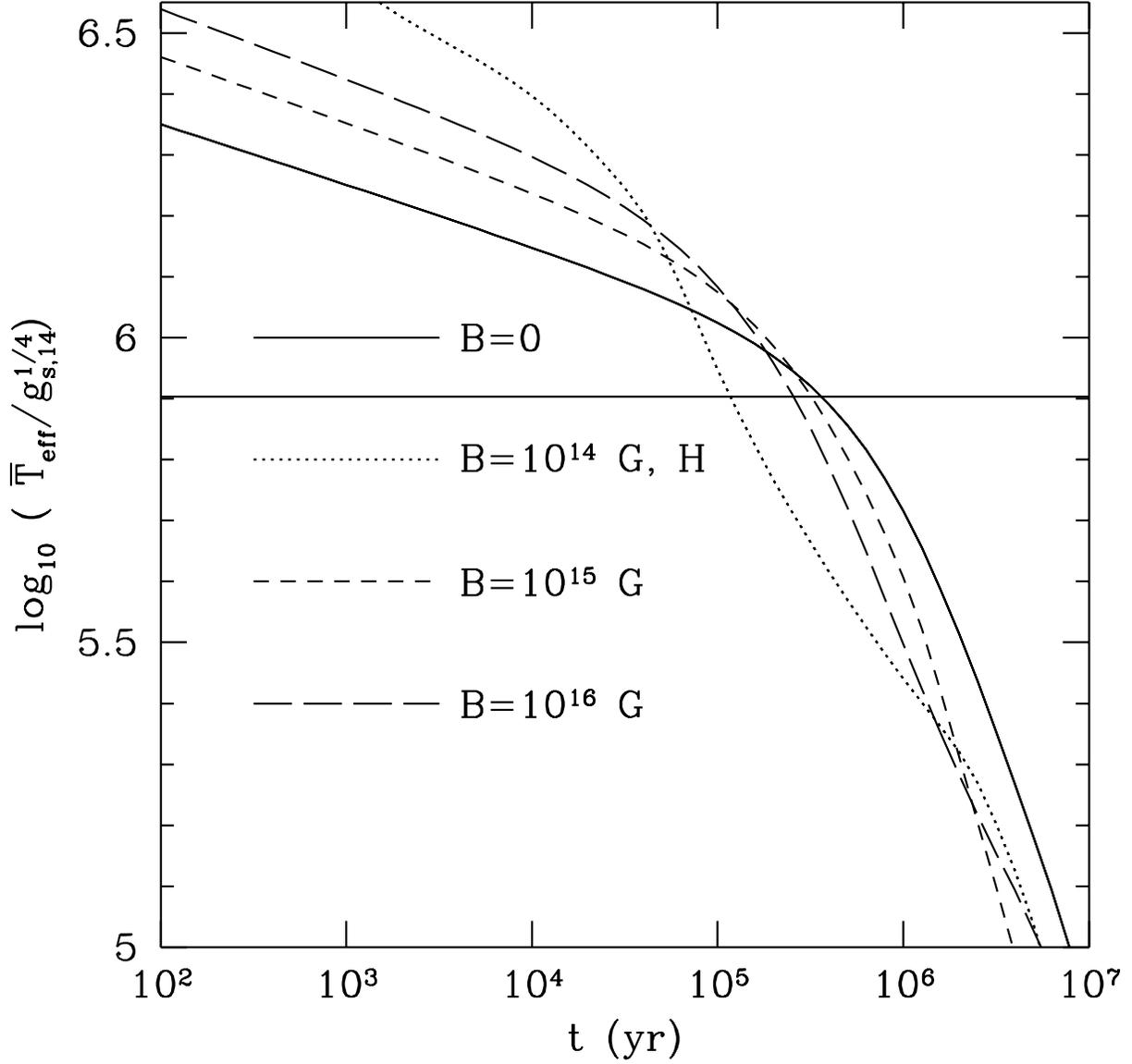}
\caption{
The curves trace the cooling evolution of isolated neutron stars by
the modified URCA process and the radiation of photons from the
surface for several field strengths with iron and hydrogen envelopes.
The horizontal line traces the
observed surface temperature of RX~J0720.4-3125 of $8 \times 10^5 K$.
}
\label{fig:rxjfig}
\end{figure}

\figref{rxjfig} depicts the thermal evolution for several values of the
magnetic field at the pole.  We have assumed that $g_s=10^{14}$ cm/s$^2$.
The horizontal line gives the value of $T_\rmscr{bb}= 8 \times 10^5$ K
found by \jcite{Habe97}.  Depending on the size and mass of the neutron
star, this line may move up or down slightly; however, this will not
strongly affect the age estimate because temperature drops quickly with
age for $t \gtrsim 3 \times 10^5$ yr.  Additionally, we see that near
this temperature, neutrino cooling and photon cooling both contribute;
consequently, the temperature at this age does not depend strongly on
field strength or composition.

From these evolutionary tracks, we find that the age of
RX~J0720.4-3125 may range from $1.2 \times 10^5$ yr for a $10^{14}$ G
field with a hydrogen envelope to $3.6 \times 10^5$ yr for an
unmagnetized neutron star.  The intense fields yield ages intermediate
to these.  We will assume a polar field of $10^{14}$ G and an iron
envelope yielding an age estimate of $3.3 \times 10^{5}$ yr.  With
this age, we can obtain an estimate of the magnetic field of the
neutron star, if we assume that its has spun down by magnetic dipole
radiation (\cite{Shap83}),
\be
t = \frac{3 I c^3}{B_p^2 R^6 \sin^2 \alpha \Omega_0^2}.
\ee
Solving for the magnetic field strength yields,
\be
B_p \sin \alpha = \sqrt{\frac{3 I c^3}{t}} \frac{P}{2\pi R^3} = 9.3 \times
10^{13} \rmmat{G} 
\left ( \frac{M}{M_\odot} \right )^{1/2} \left ( \frac{R}{10^6
\rmmat{cm} } \right )^{-2} 
\left ( \frac{t}{3.3\times 10^5\rmmat{yr}} \right )^{-1/2}
\ee
This age estimate also yields a value for $\dot P$.  For spin down by
magnetic dipole radiation, 
\be
T = \frac{P}{\dot P} = 2 t
\ee
which yields ${\dot P} = 4.3 \times 10^{-13}$.  This is consistent with
the observed ${\dot P} = -6.0 - 0.8 \times 10^{-12}$.

\section{Discussion}

We present a simple model for RX~J0720.4-3125.  Its emission
originates from neutron-star cooling through a magnetized envelope.
We derive a magnetic field $\sim 10^{14} G$, an age $\sim 3 \times
10^5$ yr and a period derivative of $4 \times 10^{-13}$.  The primary
contrast between the predictions of this model and those of
\jcite{Wang97} is the value of $\dot P$.

\jcite{Wang97} argues that RX~0720.4-3125 is an old object, $t
\gtrsim 10^9$ yr.  It initially had a much stronger field ($\sim
10^{12}$ G) than at present.  Before decaying to $\sim 10^{10}$ G,
the magnetic field spun down the neutron star to its present rate.
Wang argues that accretion powers the present emission and that a
value of ${\dot P} \lesssim 10^{-16}$ is consistent with this
field-decay model.

Our prediction of $\dot P$ lies just within the current limits
which \jcite{Habe97} obtained over a three-year baseline.  An further
round of observations within the next several years could determine
whether RX~J0720.4-3125 is a middle-aged, isolated cooling neutron
star or an ancient neutron star accreting from the interstellar
medium.

\acknowledgements

The work was supported in part by a National Science Foundation
Graduate Research Fellowship, the NSF Presidential Faculty Fellows
program and Cal Space grant CS-12-97.

\bibliography{ns}
\bibliographystyle{jer}

\end{document}